\documentclass{emulateapj}
\slugcomment{Accepted to ApJ, October 2014}

\shorttitle{Temporal Correlations in Blazars}
\shortauthors{Cohen et al.}

\begin{document}
\title{Temporal Correlations between Optical and Gamma-ray Activity in Blazars}

\author{Daniel P. Cohen\altaffilmark{1,2}, Roger W. Romani\altaffilmark{3}, Alexei V. Filippenko\altaffilmark{1}, S. Bradley Cenko\altaffilmark{4}, Benoit Lott\altaffilmark{5,6}, WeiKang Zheng\altaffilmark{1}, and Weidong Li\altaffilmark{1,7}}

\altaffiltext{1}{Department of Astronomy, University of California, Berkeley, CA 94720-3411}
\altaffiltext{2}{Department of Physics \& Astronomy, UCLA, CA 90095-1547}
\altaffiltext{3}{Department of Physics, Stanford University, Stanford, CA 94305}
\altaffiltext{4}{NASA/Goddard Space Flight Center, Greenbelt, MD 20771}
\altaffiltext{5}{Univ. Bordeaux, CENBG, UMR 5797, F-33170 Gradignan, France}
\altaffiltext{6}{CNRS, IN2P3, CENBG, UMR 5797, F-33170 Gradignan, France}
\altaffiltext{7}{Deceased 12 December 2011}

\begin{abstract}
	We have been using the 0.76-m Katzman Automatic Imaging Telescope (KAIT) at Lick Observatory 
to optically monitor a sample of 157 blazars that are bright in gamma rays, being detected with high 
significance ($\ge 10\sigma$) in one year by the Large Area Telescope 
(LAT) on the {\it Fermi Gamma-ray Space Telescope}. We attempt to observe each source on a 3-day cadence with KAIT, 
subject to weather and seasonal visibility. The gamma-ray coverage is essentially continuous. 
KAIT observations extend over much of the 5-year {\it Fermi} mission for several objects, and 
most have $>100$ optical measurements spanning the last three years. These blazars (flat-spectrum radio 
quasars and BL~Lac objects) exhibit a wide range of flaring behavior. Using the discrete correlation function 
(DCF), here we search for temporal relationships between optical and gamma-ray light curves in the 
40 brightest sources in hopes of placing constraints on blazar acceleration and emission zones. 
We find strong optical--gamma-ray correlation in many of these sources at time delays of $\sim 1$ to $\sim 10$ 
days, ranging between $-40$ and +30 days.
A stacked average DCF of the 40 sources verifies this correlation trend, with a peak above 99\% 
significance indicating a characteristic time delay consistent with 0 days. These findings strongly support 
the widely accepted leptonic models 
of blazar emission. However, we also find examples of apparently uncorrelated flares (optical flares
with no gamma-ray counterpart and gamma-ray flares with no optical counterpart) that challenge 
simple, one-zone models of blazar emission. Moreover, we find that flat-spectrum radio quasars tend to have gamma rays 
leading the optical, while intermediate and high synchrotron peak blazars with the most significant peaks have smaller lags/leads.
It is clear that long-term monitoring at high cadence is necessary to reveal the underlying
physical correlation.

\end{abstract}


\section{Introduction}

	Blazars make up a class of radio-loud active galactic nuclei (AGNs) that have a relativistic jet pointing very nearly along Earth's line of sight. These sources are generally extremely bright and highly variable from radio to gamma-ray wavebands \citep[e.g.,][]{bk79,up95}. The spectral energy distributions (SEDs) of blazars are characterized by two dominant peaks, one near radio to ultraviolet wavelengths and the other at higher, X-ray/gamma-ray energies. Optical to ultraviolet emission in blazars is widely accepted to be caused by synchrotron emission from electrons in the jet. Higher energy, hard-X-ray--GeV--TeV emission is attributed to inverse-Compton scattering (ICS) of seed photons by the synchrotron-emitting electrons (the ones responsible for the lower-energy emission), or the alternative hadronic processes based in jet proton interactions \citep[e.g.,][]{jon74,kon81,mb92}. 

In the favored leptonic models of blazar emission, synchrotron radiation and ICS both occur along the jet and derive from the same population of electrons, yielding a strong correlation between low- and high-energy wavebands.   \citep[e.g.,][]{sik94}.  Observed flares derive from events, commonly modeled as propagating shocks (or collisions of shocks), that occur in the jet at subparsec to parsec distances from the central engine and accelerate the jet electrons to high energies \citep[e.g.,][]{spa01}. While synchrotron photons are emitted near the shock front in the jet, the origin of the seed photons for ICS is not clear. These seed photons could be produced in the synchrotron-emitting jet itself (synchrotron self-Compton [SSC]) or from an external source (external Compton [EC]) such as the accretion disk, broad-line region (BLR), or dusty infrared torus (hot-dust region [HDR]) \citep[e.g.,][]{jon74,sik94}.  

	Multi-wavelength correlation studies of blazars can thus help to place constraints on the dominant mechanisms driving variability and identify the relationship between emission zones. For example, the leptonic models predict a strong correlation between synchrotron-produced optical and ICS-produced gamma-ray emission. Lags or leads of high significance between flares in these wavebands may help place constraints on the location of the ICS seed photons relative to the synchrotron-emitting shock in the jet and discern between the SSC and EC processes. Alternatively, observations of a flare in one waveband with no correlated flare in the other might suggest multiple zones of emission or support hadronic models of blazar emission. With the advent and success of the {\it Fermi Gamma-ray Space Telescope} and its primary scientific instrument the Large Area Telescope ({\it Fermi}/LAT), multi-wavelength studies have been extended into this MeV--GeV energy range and these goals are being realized.  For example, \citet{fur14} and \citet{max14} both present fascinating investigations of radio--gamma-ray correlations in blazars utilizing the discrete correlation function (DCF) -- the former using cm to sub-mm data from the F-GAMMA monitoring project and the latter using 15~GHz data from the Owens Valley 40-m telescope.
	
        In this study, we use the DCF to investigate correlations between optical and gamma-ray light curves of blazars. Optical data were collected with the robotic 0.76-m Katzman Automatic Imaging Telescope (KAIT) at Lick Observatory, which has been monitoring sources detected by LAT for much of the {\it Fermi} mission. Here we present results from computing the DCFs between optical and gamma-ray light curves in the 40 brightest sources out of the 157 monitored blazars. A future paper will report the results for the other sources.

        This paper is organized as follows. In \S \ref{obs} we describe data collection and production of the light curves. Section \ref{lags} and \S \ref{stacks} present the results and interpretation for DCFs of individual sources and for stacked DCFs of subsets of sources, respectively. We conclude in \S \ref{disc} with a brief discussion of our findings. 

\section{Observations} \label{obs}
\subsection{KAIT} 

	Since August 2009, we have been using KAIT to monitor gamma-ray bright blazars. The base sample consists of the blazars at Galactic latitude $|b|>10^{\circ}$ in the KAIT declination band $-25^{\circ}<\delta<70^{\circ}$ detected in the first-year LAT blazar catalog \citep{1LAC} at a significance $>10\sigma$ and in the historical optical (POSS) at $R<18$ mag. There were 140 such sources. A few additional optical/gamma-bright blazars have been added to bring the monitored sample to 157. Every available night we attempt unfiltered observations, with effective color close to that of the $R$ band \citep{li03}, of 30--50 sources. Light curves are produced through a pipeline that utilizes aperture photometry and performs brightness calibrations using USNO B1.0 catalog stars in each source field. Currently, KAIT light curves typically contain at least 100 data points with an average in-season cadence of $\sim 3$ days, extending over much of the full continuous 5-year {\it Fermi} coverage. All KAIT AGN light curves are made publicly available on the web at \url{http://hercules.berkeley.edu/kait-agn} and are updated in nearly real time.
	
\subsection{{\it Fermi}/LAT}

	The {\it Fermi}/LAT data were collected over the first 63 months of the mission from 2008 August 4 to 2013 November 4. Time intervals during which the rocking angle of the LAT was greater than 52$^{\circ}$ were excluded and a cut on the zenith angle of gamma rays of $100^{\circ}$ was applied. The Pass 7\_V6 Clean event class was used, with photon energies between 100 MeV and 200 GeV.

	The LAT light curves were produced from variable-width bins, generated by the adaptive binning method (ABM), to obtain nearly constant relative-flux uncertainties of 25\% in each bin \citep{lott12}. The systematic uncertainties are negligible relative to the statistical uncertainties in these light curves. In the following, we use the ``optimum energy,'' defined by \citet{lott12} as the lower limit of the integral fluxes shown in the light curves. For the optimum energy, the accumulation times (i.e., bin widths) needed to fulfill the condition on the relative-flux uncertainty with the ABM are the shortest (on average) relative to other choices of lower energy limit. Because the sources are variable and the optimum energy value depends on the flux, we compute the optimum energy with the average flux over the first two years of LAT operation reported in the 2FGL catalog \citep{2FGL}.

	The photon index was set fixed to the value reported in the 2FGL catalog when assessing the time intervals with the ABM.  For heavily confused sources, the fluxes of the main (maximum 3) neighboring sources were fitted as well in the ABM. Once the time intervals were determined with the ABM,  the final analysis  was performed with the unbinned likelihood method implemented in the  {\it pyLikelihood} library  of the Science Tools\footnote{http://fermi.gsfc.nasa.gov/ssc/data/analysis/\\documentation/Cicerone/} (v9r32p5). The spectra were modeled with single power-law functions, with both fluxes and photon indices set free for the source of interest as well as the sources found variable in the 2LAC \citep{2LAC} and located within $10^\circ$ of the source of interest. 
	
	Thus far, ABM light curves with coverage until the end of June 2013 have been produced for the 40 brightest sources in our sample of 157 blazars, selected according to detection significance after 2 years of {\it Fermi}/LAT operation.
	 These sources are listed in Table \ref{tab} with optical classification, SED classification, and redshift obtained from the 2LAC along with the optimum energy for each source. Also given is the duration of overlap for the KAIT and {\it Fermi}/LAT light curves --- that is, the fraction of the LAT months containing at least one KAIT observation. Our sample includes 20 BL~Lac objects and 20 flat-spectrum radio quasars (FSRQs); for the BL~Lacs there are 6 low synchrotron peak (LSP), 8 intermediate synchrotron peak (ISP), and 6 high synchrotron peak (HSP) as defined in the 2LAC.

\section{Individual Time Lags and Significances, Flaring Behaviors} \label{lags}

	We compute the DCFs \citep{ek88}, using the local normalization as in \citet{wel99}, between the KAIT and {\it Fermi}/LAT light curves for the blazars listed in Table \ref{tab}. With the local normalization, the points in the DCF are bounded between $-1$ and 1, and they straightforwardly represent the linear correlation coefficient for each lag bin (see \citealt{whi94} for more on this technique). Levels of significance for each source are derived from the distribution of DCF points in each lag bin for false-match source pairs: we compute the DCF of the {\it Fermi}/LAT light curve of one source with the KAIT light curve of a {\it different} source for all possible false-pair matches (typically a few thousand pairs) using the 40 ${\it Fermi}$/LAT and full 157 KAIT light curves, yielding a distribution of correlation points at each lag bin. For this calculation, we use KAIT light curves within a right ascension of $\pm 3$ hr of the actual source of interest (thereby using sources with similar optical coverage to the actual source of interest). The levels of significance calculated are then expected to contain uncertainties from systematics, uncorrelated flaring events, and possibly window functions of the light curves. 

        In all plots showing a DCF in this paper, dashed blue, green, and orange lines represent 68\%, 90\%, and 99\% significance levels, respectively. Centroid lags and uncertainties are derived from weighted least-squares Gaussian fits to DCF points in the peak. Peak significance values and errors are also derived from the Gaussian fits, converting the amplitude values to probability values using the false-match correlation distribution. Individual DCF points are occasionally higher, but within the individual point measurement errors. Wide DCF peaks may indicate either a range of characteristic timescales in the correlated response, or simple measurement limitations. Note that there may be unstudied correlations in the bins near the DCF peaks, especially since the adaptive binning light curves produce large time bins in low states which may spread the effective correlation among a range of lags. However, since the DCF peaks are dominated by LAT flaring events, we believe that this does not cause undue smoothing when the sources are relatively bright and the time bins are small.

\begin{figure*}
\centering
\includegraphics[scale=0.7]{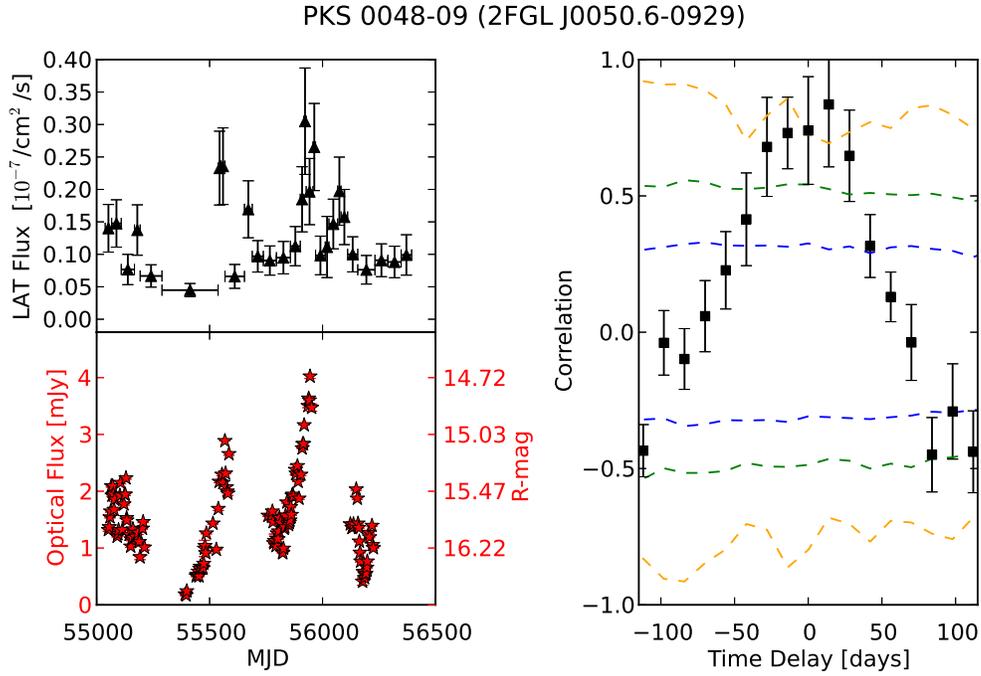}
\caption{The {\it Fermi}/LAT light curve (top left), KAIT light curve (bottom left), and DCF (right) for the BL~Lac PKS~0048$-$09. Blue, green, and orange dashed lines on the DCF plot represent 68\%, 90\%, and 99\% significance levels, respectively. This source has optical and gamma rays strongly correlated at $\tau = -5.3 \pm 3.1$ days with a peak DCF at 99.0\% significance. \label{0050}}
\end{figure*}

\begin{figure*}
\centering
\includegraphics[scale=0.7]{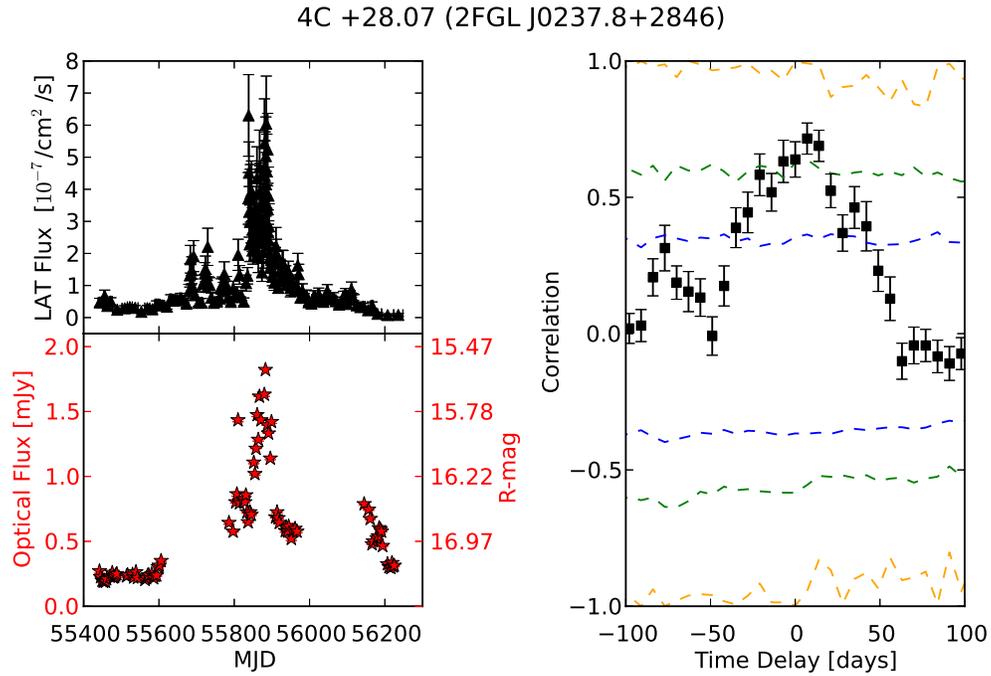}
\caption{FSRQ 4C~+28.07, with a DCF peak (90.8\% significance) at $\tau = 4.1 \pm 1.3$ days. As in Figure 1.\label{0237}}
\end{figure*}

\begin{figure*}
\centering
\includegraphics[scale=0.7]{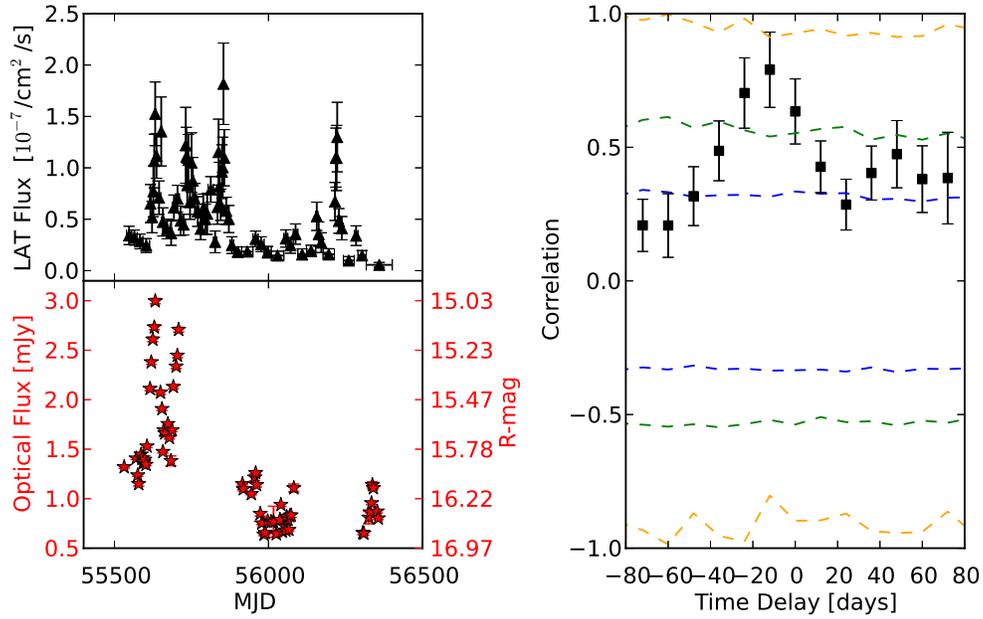}
\caption{BL~Lac 4C~+01.28, with a DCF peak (97.7\% significance)  at $\tau = -13.8 \pm 3.1$ days. As in Figure 1.\label{1058}}
\end{figure*}

\begin{figure*}
\centering
\includegraphics[scale=0.7]{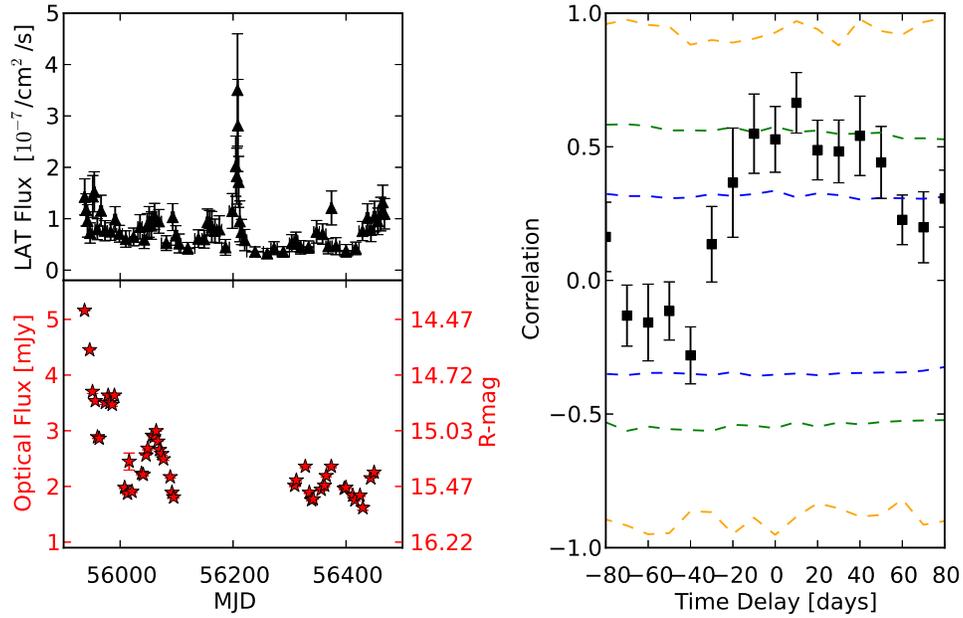}
\caption{FSRQ 3C~279, with a DCF peak (92.3\% significance) at $\tau = 19.7 \pm 3.4$ days. Unfortunately, optical coverage missed 
a very large gamma-ray flare at MJD $\approx$ 56,200, so the correlation here is dominated by smaller-scale variability. As in Figure 1.\label{1256}}
\end{figure*}

\begin{figure*}
\centering
\includegraphics[scale=0.7]{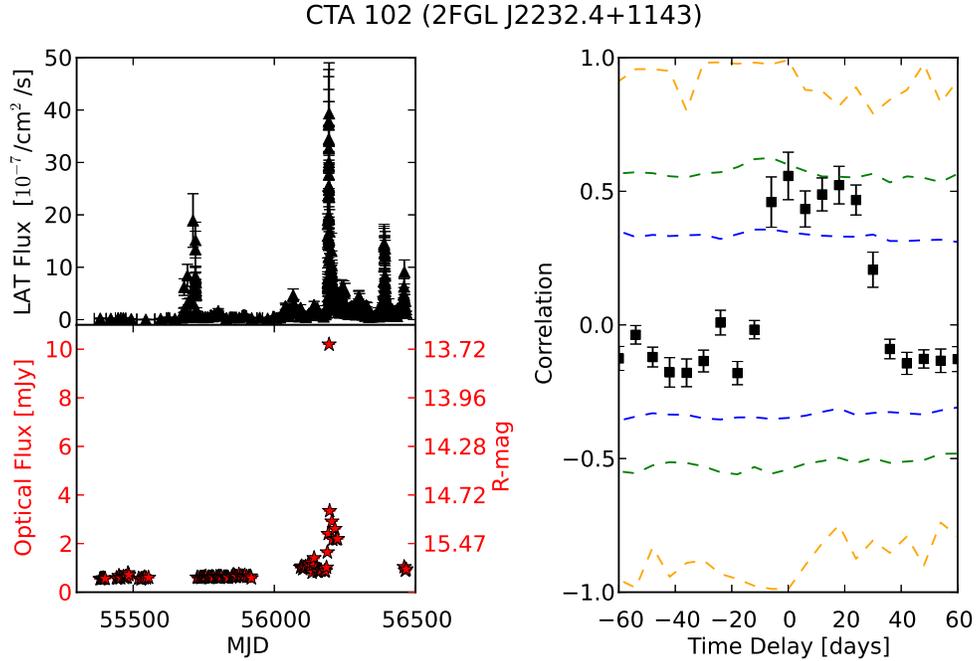}
\caption{FSRQ CTA 102, with a DCF peak (92.3\% significance) at $\tau =11.4\pm0.7$ days, dominated by the large flare 
near MJD $\approx$ 56,200. As in Figure 1.\label{2232}}
\end{figure*}

	The time delay in all DCF plots (and throughout this paper) is defined so that positive time delay $\tau > 0$ corresponds to gamma rays leading optical emission. Peak significance values and centroid lags derived from the DCF for each source are given in Table \ref{tab}. For some of the sources the DCF was either flat or highly scattered, with no clear peak to fit; such centroid lags and peak significances are marked with ``---'' in Table \ref{tab}. Examples of DCFs are shown in Figures \ref{0050}--\ref{2232}. 

	From the light curves, perhaps the clearest correlation among these sources (although not the most significant) is in the FSRQ 4C~+28.07, with nearly simultaneous gamma-ray and optical flares at MJD $\approx$ 55,900. The DCF peak for 4C~+28.07 indicates a gamma-ray lead of $\tau = 4.1 \pm 1.3$ days. Visually, the good correlation in the brightest flares supports this short time scale, but the DCF peak width at half maximum is $\sim 80$ days. This likely represents the characteristic width of the flaring episodes, but may also describe a variation in the true lead/lag. Long time series with roughly 1-day sampling would be needed to distinguish these cases. Nevertheless, our fit time scale does provide an estimate for the typical lead/lag time, indicating relatively tight optical--gamma-ray correlation. Similar considerations apply to CTA 102, another FSRQ, where we see nearly simultaneous gamma-ray and optical flares at MJD $\approx$ 56,200. This results in a correlation peak indicating a gamma-ray lead of $\tau= 11.4 \pm 0.7$ days. The strong dominant peak limits the range of $\tau$ contributions to $-10$ to +25 days. The gamma-ray lead of the FSRQ 3C~279 by $19.7\pm 3.4$ days is less clear from its light curve, as the correlation peak in this source derives from somewhat smaller-scale variability (no large flares with overlapping optical and gamma-ray coverage). The BL~Lacs PKS~0048$-$09 and 4C~+01.28 are strongly correlated at $\tau = -5.3 \pm 3.1$ and $\tau = -13.8 \pm 3.1$ days, respectively. PKS~0048$-$09 owes its correlation to a pair of correlated optical and gamma-ray flares at MJD $\approx$ 55,500 and MJD $\approx$ 55,800, and 4C~+01.28 to an early flare at MJD $\approx$ 55,600.
	
	A few of the sources in our sample have been the subjects of other multi-wavelength studies. Most recently, in \citet{ack14}, the DCF for the FSRQ 4C~+21.35 suggested gamma rays leading the optical by $\sim 35$ days during a flare in 2010, while the DCF we computed indicates a gamma-ray lead of $\tau = 8.6 \pm 1.5$ days. \citet{hay12} studied 3C~279 and found gamma rays to lead the optical by $\sim 10$ days for a flare just before KAIT coverage began. The FSRQ 3C~454.3 was shown by \citet{bon09} to be correlated at $\sim 0$ days for a flare that occurred before KAIT coverage began; unfortunately, for this source there is no overlap between the KAIT and {\it Fermi}/LAT light curves, so the DCF could not be computed. Discrepancies in DCFs between this and other studies of the same sources are caused by different observations of different flaring activity.  
	
\begin{figure*}
\centering
\includegraphics[scale=0.6]{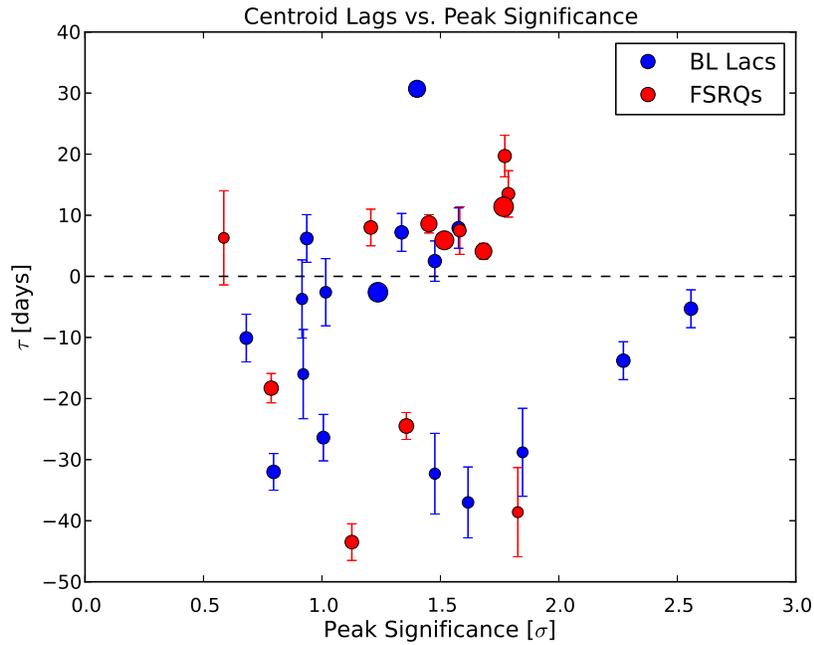}
\caption{Scatter plot of centroid lags/uncertainties and peak significances for the sources in Table \ref{tab} with measured DCF peaks. Data points with small lag uncertainty are shown as larger symbols, for clarity.  \label{scatter}}
\end{figure*}

	Of the 40 sources in our study, 8 are found to have DCF peaks above 90\% significance. We find in general that these sources have strong optical--gamma-ray correlation with timescales on the order of days to tens of days, in agreement with similar studies of other sources \citep[e.g.,][]{abd10a,abd10c,ack12,hay12,bon09,bon12}. In order to further visualize and understand the distribution of lags, we include a scatter plot in Figure \ref{scatter} that shows the centroid lags and uncertainties versus the peak significances (in terms of Gaussian probability $\sigma$). For clarity, data points with high certainty of centroid lag value are shown to be larger than those with uncertain lags. The FSRQs, with several exceptions, tend to have gamma rays leading the optical by 0--20 days, while the BL~Lacs are widely scattered with no clear trend toward lead or lag. This behavior apparently supports the  current models that suggest EC is dominant in FSRQs while SSC is dominant in BL~Lacs \citep[e.g.,][]{bot13}. However, with a small sample size this result should be treated with caution.

	Models for SSC and those for EC do predict time delays between the optical and gamma-ray bands. For EC, the HDR and BLR are predicted to dominate the contribution from external radiation fields \citep{sik09}. Under the assumption that flares are produced by outbursts propagating down the jet, optical--gamma-ray leads and lags of roughly day timescales are predicted by the strong stratification of the radiation field, and its mismatch with the decreasing magnetic energy of the jet. Applying the EC model presented by \citet{jan12}, the preference of FSRQs (dominated by EC) to have gamma rays leading the optical by these timescales suggests that, for FSRQs, the locations of source activity (i.e., flare burst events) occur more often downstream but still within the radiation field of the BLR, or well downstream but still within the radiation field of the HDR. SSC models predict time delays, again of both signs, but generally of smaller magnitude \citep[e.g.,][]{sok04}. With strong optical--gamma-ray correlations for many sources, our results support leptonic single-zone models (both SSC and EC) of blazar emission; in the hadronic models, the low- and high-energy SED peaks vary independently, and strong correlation between optical and gamma-ray emission is not expected. 

\subsection{Uncorrelated Flares} \label{uncflares}

\begin{figure*}
\centering
\includegraphics[scale=0.6]{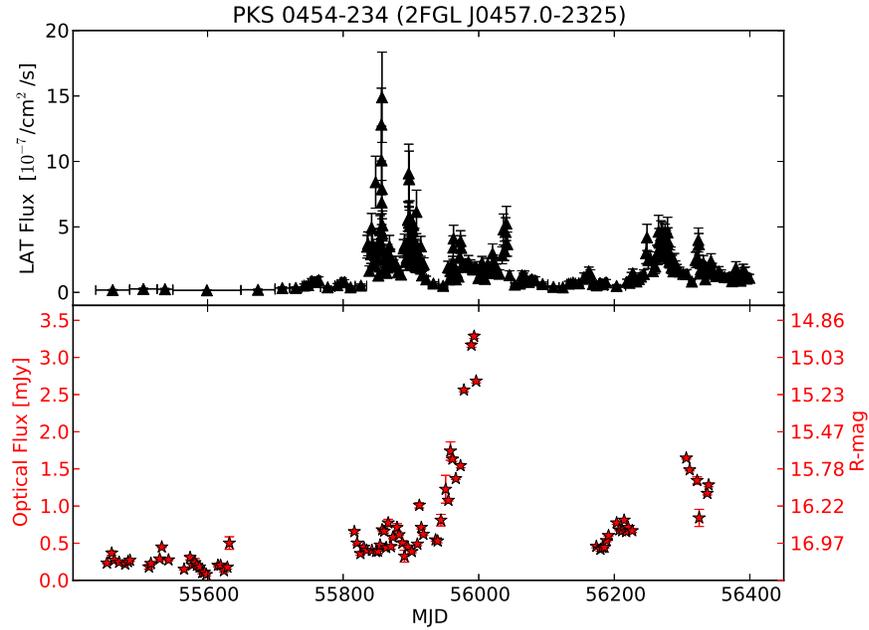}
\caption{The {\it Fermi}/LAT light curve (top) and KAIT light curve (bottom) of FSRQ PKS~0454$-$234, showing an example of a large gamma-ray flare with no (or very small) optical counterpart, at MJD $\approx$ 55,850. \label{0457}}
\end{figure*}

	There are a few sources among our 40 that exhibit peculiar flaring behavior --- large gamma-ray 
flares with little or no optical counterpart or optical flares with no gamma-ray counterpart. The 
former are commonly called ``orphan'' flares, and have been attributed to hadronic processes 
\citep[e.g.,][]{bot07} or alternatively to contamination in the optical by accretion-disk emission \citep{ack14}, but the origin of these flares remains uncertain. 
We see several such flares, such as in the light curve of the FSRQ PKS~0454$-$234 in Figure \ref{0457}, 
at MJD $\approx$ 55,850, where several strong gamma-ray flares show little optical activity nearby in time.
The next large optical flare peaks at MJD $\approx$ 56,000; if it is the counterpart of the gamma-ray
activity, it suggests a unique $\sim 150$ day delay (the alternative is that this optical flare is also an orphan). A similar large delay between X-rays/gamma-rays was detected (in the DCF) for 3C~279 by \citet{hay12}; however, a causal connection was ruled out owing to the lack of accompanying radio flares and the temporal structure of the X-ray and gamma-ray flares. In our case of PKS~0454$-$234 and others like it, substantially longer light curves or more multi-wavelength data are needed to confirm or dismiss a causal origin of such large delays. 

\begin{figure*}
\centering
\includegraphics[scale=0.6]{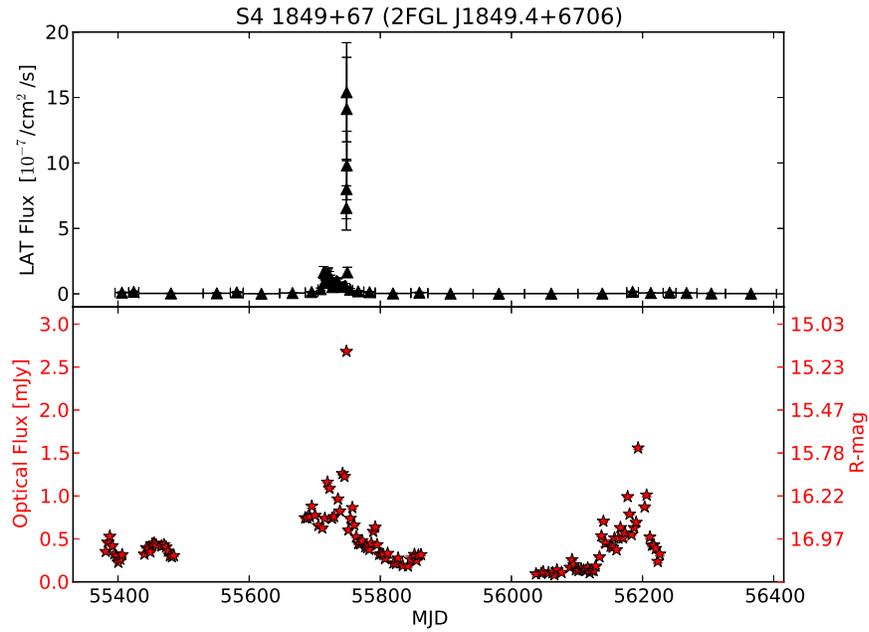}
\caption{The {\it Fermi}/LAT light curve (top) and KAIT light curve (bottom) of S4~1849+67, a FSRQ,  exhibits complex synchrotron variability with an optical flare that is correlated with gamma rays at MJD $\approx$ 55,750, and a second broad optical flare at MJD $\approx$ 56,150 with no gamma-ray counterpart. \label{1849}}
\end{figure*}

	In the sources BZQ~J0850$-$1213, OP~313, and S4~1849+67, we also observe clear optical variability 
with no correlated gamma-ray activity. For the FSRQ S4~1849+67, shown in Figure \ref{1849}, the 
nearly simultaneous optical and gamma-ray peaks at MJD $\approx$ 55,750 indicate a small characteristic
delay; the lack of gamma-ray activity near the second broad optical flare at MJD $\approx$ 56,100 
is thus particularly significant.  Optical flares with no gamma-ray counterpart have also been noted
by other authors in various sources \citep[e.g.,][]{smi11,dam13}. Such behavior has been attributed 
to separate emission zones for the flare outbursts or synchrotron emission modulation via
magnetic-field changes, with minimal effect on the seed photon numbers, and thus relatively
steady EC emission \citep[e.g.,][]{cha13}. Thus, orphan events argue for multi-zone synchrotron
sites, with decoupled Compton emission.
	
\section{Stacked Correlations} \label{stacks}

	In order to assess correlation trends between optical and gamma-ray emission in 
blazars, we stack, or average, DCFs for our 40 sources and for subsets of this sample. Significance 
levels for stacked DCFs are again derived from false-match correlations, but now each point in 
the false-match distribution is an average, using the same subclass, for as many sources as are used in the true stacked DCF. For example, if there are 20 FSRQs into the stacked DCF, then each point in the false-match distribution is an average of 20 false-match pair correlations, where the false-match pairs are drawn from subsets of only FSRQs. There are typically $\sim 100$ points in the false-match distribution in any given bin that are then used to calculate significance levels for stacked sources (compared with a few thousand for individual source false-match distributions). 

\begin{figure*}
\centering
\includegraphics[scale=0.6]{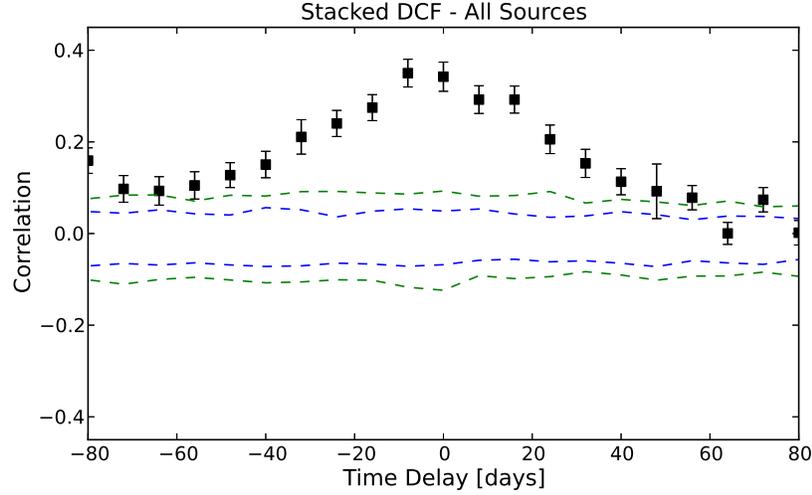}
\caption{The average, or stacked, DCF for all 40 sources with 68\% (blue) and 90\% (green) significance levels. The peak is centered at $\tau = -2.8 \pm 1.8$ days and well above all points in the stacked false-match distribution, indicating a significance above 99\%. The lag and its uncertainty are calculated here by again fitting a Gaussian to the DCF peak. \label{stackfull}}
\end{figure*}

      The stacked DCF for the full set of 40 sources is shown in Figure \ref{stackfull}. The peak 
is much higher than any one of the $\sim 100$ stacked false-pair points, indicating a significance above 99\%. The centroid lag is $\tau = -2.8 \pm 1.8$ days (the lag value and its uncertainty for the stacked DCFs are again determined through a Gaussian fit). This very strong 
correlation signal indicates that, on average, blazars have strongly coupled optical--gamma-ray 
emission with a characteristic lag of roughly 1 day, most easily understood for
leptonic models. The peak width implies apparently significant correlation from about $-60$ to +40 days, consistent with the range of characteristic lags from individual DCFs (from about $-40$ to +30 days). This dispersion may alternatively be dominated by the typical duration of the flare events. Indeed, we find that this stacked DCF is not much broader than the best individual DCFs, suggesting that $|\tau| \approx 1$ day delays are typical of our monitored sample. The peak fit indicates a small optical lead, which might be caused by typically larger (and more uncertain) characteristic time delays found for optical leads while significant gamma-ray leads tend to cluster around smaller values of $\tau$ (Fig. \ref{scatter}). The $\tau < 0$ day characteristic lag for the stacked DCF peak might also reflect larger characteristic widths for the optical events: longer optical fall times would bias the correlation overlaps to negative $\tau$, especially for weaker flares.

\begin{figure*}
\centering
\includegraphics[scale=0.6]{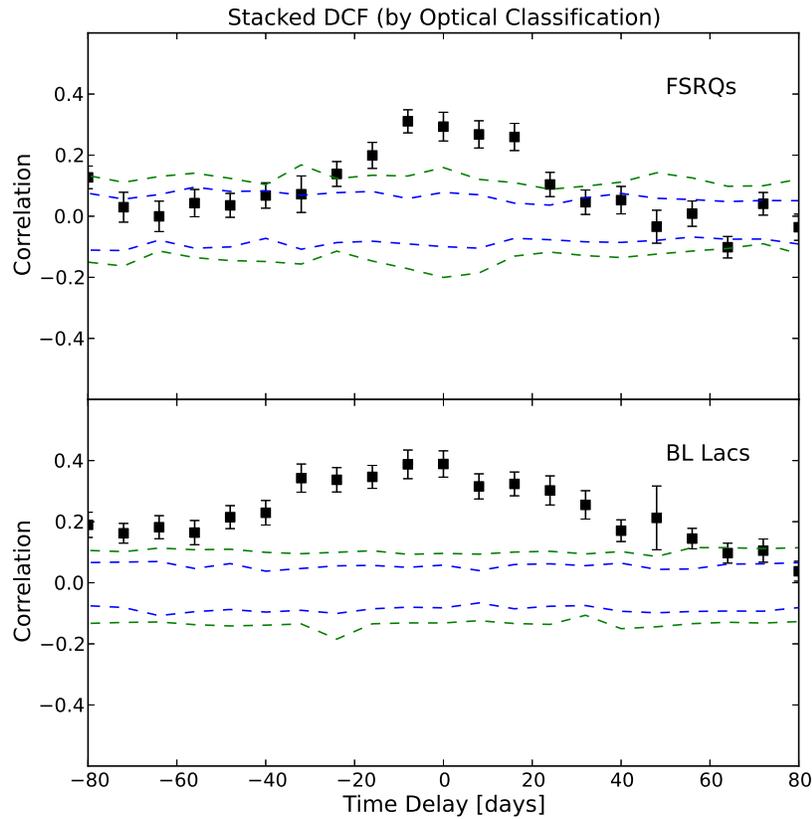}
\caption{Stacked DCFs for the 20 FSRQs and 20 BL~Lacs in our sample with 68\% (blue) and 90\% (green) significance levels. Both peaks are above any one point in the stacked false-match distributions, indicating $>99$\% significance. The DCFs indicate centroid lags at $\tau = -0.6 \pm 2.1$ and $\tau = -5.6 \pm 3.5$ days for FSRQs and BL~Lacs, respectively. \label{stackopt}}
\end{figure*}

	The stacked DCFs for the 20 FSRQs and 20 BL~Lacs in our sample are shown in Figure \ref{stackopt}. 
There are a few interesting features to note from these average correlations. First, although both 
peaks are well above any one of the stacked false-pair distribution points, the peak for FSRQs appears less significant than that for 
BL~Lacs. If BL~Lacs are indeed dominated by SSC and FSRQs by EC, then this trend is expected, as the correlations for EC are somewhat weaker than for SSC (quadratic for SSC and linear for EC). Second, the stacked DCF 
peak for FSRQs is a factor of $\sim 2$ narrower than that for BL~Lacs. The relative tightness of this stacked correlation peak is likely caused by FSRQs having
higher typical variability than BL~Lacs \citep[e.g.,][]{abd10d,2LAC}, on average leading to better determined correlation peaks and more tightly constrained lags. The larger spread in BL~Lacs may then be caused by a wider distribution of effective lags between the sources or a lower amplitude and longer timescale for the typical flaring event. If the significance of individual source DCF peaks were to improve with better coverage of overlapping flares, we might expect the combined BL~Lac DCF to settle down
to the mean lag expected for SSC emission.

\begin{figure*}
\centering
\includegraphics[scale=0.6]{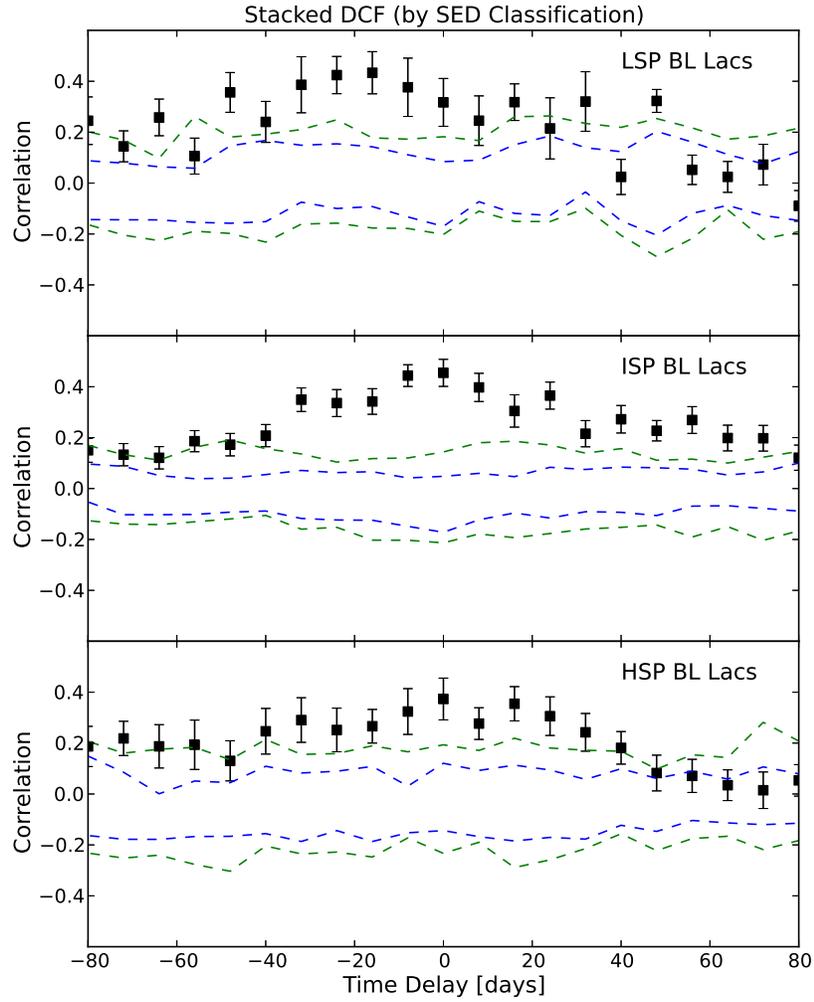}
\caption{Stacked DCFs for the 6 LSP, 8 ISP, and 6 HSP BL~Lacs in our sample of 40 sources. Blue and green dashed lines represent 68\% and 90\% significance levels, respectively. While the maximum DCF bins are significant, the fits to the DCF peaks are quite uncertain, indicating
inadequate measurement or a wide range of characteristic lags. The fits show lags at $\tau = -17.4 \pm 21.1$, $\tau = -1.6 \pm 3.9$, and $\tau = -6.0 \pm 6.7$ days for LSP, ISP, and HSP BL~Lacs, respectively. \label{stacksed}}
\end{figure*}

	The location of the synchrotron peak in the SEDs might also correlate with DCF peak properties.
FSRQs are LSP blazars, while BL~Lacs have a wider distribution of synchrotron peak energy, including ISP 
and HSP BL~Lacs. One might expect the latter to be increasingly SSC dominated.
The stacked DCFs for the 6 LSP, 8 ISP, and 6 HSP BL~Lacs are shown in Figure \ref{stacksed}. At present, we have 
too few objects to infer real trends; the ISP BL Lacs exhibit the only well-defined peak, at $\tau = -1.6 \pm 3.9$ day.

\section{Discussion} \label{disc}

	We have been monitoring 157 gamma-ray-bright sources detected by {\it Fermi}/LAT with KAIT at Lick Observatory, and here present a study of optical--gamma-ray correlations in 40 sources selected based on {\it Fermi}/LAT detection significance. Overall, optical and gamma-ray emission were found to be highly correlated in these sources, at time delays of roughly days. An average DCF for all 40 sources was found to have a peak significance above 99\%, confirming the strength in correlation between the two wavebands. Such strong correlations support the leptonic models of ICS gamma-ray emission, in which seed photons are upscattered by relativistic jet electrons responsible for synchrotron optical emission. 

        Whether the seed photons are dominated by the synchrotron radiation produced by these electrons (SSC) or by external radiation (EC) is difficult to discern based on lags and correlation strengths of these sources. However, we find that the well-measured FSRQs tend to have positive lags (gamma rays leading the optical) while the best-measured BL~Lacs show no clear trend toward lag or lead. This supports models with EC being dominant in FSRQs and SSC dominant in BL~Lacs. Stacked DCFs of LSP, ISP, and HSP BL~Lacs are consistent with increasing SSC dominance as synchrotron peak energy increases; ISP and HSP BL~Lacs are found to have average DCF peak lags closer to 0 days than LSP BL~Lacs. However, a larger sample is required to make definitive claims. We plan on performing a similar study with {\it Fermi}/LAT light curves for the full set of 157 blazars being monitored in order to verify these findings based on the 40 brightest sources.
	
	  Recently, optical--gamma-ray correlations in blazars have been investigated through modulation indices (rather than DCFs and lags between wavebands) by \citet{hov14}. With a very large sample size, this study found that HSP BL~Lacs were most strongly and tightly correlated, supporting the notion that SSC becomes more prevalent in ISP and HSP sources while EC is more dominant in LSP sources. Based on the 40 sources in our study, we also find a stronger average correlation in BL~Lacs (but narrower average peak for FSRQs, likely owing to the high variability of FSRQs in our sample). Observationally, this trend is somewhat surprising, as the strongest correlations in individual sources have been found for FSRQs \citep[e.g.,][]{hay12}. That BL~Lacs are on average more strongly correlated than FSRQs will need further testing, although our findings and those of \citet{hov14} support this claim. 
	
	We have shown that strongly flaring, well-measured FSRQs tend to show gamma-ray leads, supporting the conclusions of other variability studies.  However, the situation for the BL~Lacs is evidently more complex. Multi-year, multi-wavelength monitoring at high cadence is clearly needed to probe the mechanisms driving variability in these sources and possible connections to the central engines. With further KAIT coverage and the continued success of {\it Fermi}/LAT, we can start to break down this complexity and understand these objects at their most fundamental level. 
		
\acknowledgments

The \textit{Fermi} LAT Collaboration acknowledges generous ongoing support from a number of agencies and institutes that have supported both the development and the operation of the LAT as well as scientific data analysis.  These include the National Aeronautics and Space Administration (NASA) and the Department of Energy in the United States, the Commissariat \`a l'Energie Atomique and the Centre National de la Recherche Scientifique/Institut National de Physique Nucl\'eaire et de Physique des Particules in France, the Agenzia Spaziale Italiana and the Istituto Nazionale di Fisica Nucleare in Italy, the Ministry of Education, Culture, Sports, Science, and Technology (MEXT), High Energy Accelerator Research Organization (KEK), and Japan Aerospace Exploration Agency (JAXA) in Japan, and the K.~A.~Wallenberg Foundation, the Swedish Research Council, and the Swedish National Space Board in Sweden. Additional support for science analysis during the operations phase is gratefully acknowledged from the Istituto Nazionale di Astrofisica in Italy and the Centre National d'\'Etudes Spatiales in France.

The work was financed in part by NASA grants NNX10AU09G, GO-31089, NNX12AF12G, and NAS5-00147. We are also grateful for support from Gary and Cynthia Bengier, the Richard and Rhoda Goldman Fund, the Christopher R. Redlich Fund, the TABASGO Foundation, and NSF grant AST-1211916.  KAIT and its ongoing operation were made possible by donations from Sun Microsystems, Inc., the Hewlett-Packard Company, AutoScope Corporation, Lick Observatory, the NSF, the University of California, the Sylvia and Jim Katzman Foundation, and the TABASGO Foundation.  We dedicate this paper to the memory of our dear friend and collaborator, Weidong Li, whose unfailing devotion to KAIT was of pivotal importance for this work; his premature, tragic passing has deeply saddened us.

Facilities: \facility{Lick:KAIT}, \facility{Fermi:LAT}


\clearpage
\begin{deluxetable}{llccccccc}
\tabletypesize{\scriptsize}
\tablewidth{\textwidth}
\tablecaption{\label{tab} Properties of blazars in our sample, with approximate KAIT--{\it Fermi}/LAT overlap duration}
\centering
\tablehead{
\colhead{Source Name} & \colhead{2FGL} & \colhead{Optical Class} &
\colhead{SED Class} & \colhead{$z$} & \colhead{$E_{\rm opt}$ [MeV]} & \colhead{Lag [days]} &
\colhead{Peak Sig. \%} & \colhead{Overlap [\%]} } 
\startdata

PKS 0048$-$09 & J0050.6$-$0929 & BL Lac & ISP & 0.635 & 313 & $-5.3\pm3.1$ & $99.0_{-1.7}^{+0.7}$ & 42/54 \\
S2 0109+22 & J0112.1+2245 & BL Lac & ISP & 0.265 & 297 & $7.2\pm3.1$ & $81.8_{-6.7}^{+5.8}$ & 32/54 \\
B3 0133+388 & J0136.5+3905 & BL Lac & HSP & --- & 546 & --- & --- & 33/54 \\
3C 66A & J0222.6+4302 & BL Lac & ISP & --- & 293 & --- & --- & 31/54 \\
4C +28.07 & J0237.8+2846 & FSRQ & LSP & 1.206 & 283 & $4.1\pm1.3$ & $90.8_{-1.8}^{+1.5}$ & 27/48 \\
PKS 0301$-$243 & J0303.4$-$2407 & BL Lac & HSP & 0.260 & 337 & $7.9\pm3.3$ & $88.6_{-6.7}^{+5.2}$ & 31/62 \\
PKS 0420$-$01 & J0423.2$-$0120 & FSRQ & LSP & 0.916 & 235 & $7.5\pm3.9$ & $88.6_{-6.4}^{+5.4}$ & 32/54 \\
PKS 0454$-$234 & J0457.0$-$2325 & FSRQ & - & 1.003 & 220 & --- & --- & 32/54 \\
4C +14.23 & J0725.3+1426 & FSRQ & LSP & 1.038 & 293 & $13.5\pm3.8$ & $92.6_{-4.3}^{+1.9}$ & 23/50 \\
PKS 0805$-$07 & J0808.2$-$0750 & FSRQ & LSP & 1.837 & 322 & --- & --- & 27/53 \\
PKS 0829+046 & J0831.9+0429 & BL Lac & LSP & 0.174 & 319 & $-28.8\pm7.2$ & $93.5_{-7.5}^{+3.5}$ & 29/53 \\
BZQ J0850$-$1213 & J0850.2$-$1212 & FSRQ & LSP & 0.566 & 335 & --- & --- & 28/56 \\
S4 0917+44 & J0920.9+4441 & FSRQ & LSP & 2.189 & 237 & $8.0\pm3.0$ & $77.2_{-10.9}^{+8.0}$ & 33/62 \\
4C +55.17 & J0957.7+5522 & FSRQ & LSP & 0.899 & 320 & --- & --- & 31/56 \\
1H 1013+498 & J1015.1+4925 & BL Lac & HSP & 0.212 & 390 & $-2.6\pm5.5$ & $69.1_{-8.4}^{+5.9}$ & 32/56 \\
4C +01.28 & J1058.4+0133 & BL Lac & LSP & 0.888 & 264 & $-13.8\pm3.1$ & $97.7_{-2.4}^{+0.8}$ & 30/57 \\
TXS 1055+567 & J1058.6+5628 & BL Lac & ISP & 0.143 & 367 & $2.5\pm3.3$ & $86.1_{-10.4}^{+6.9}$ & 33/55 \\
Ton 599 & J1159.5+2914 & FSRQ & LSP & 0.725 & 225 & $-43.5\pm3.0$ & $74.0_{-7.1}^{+6.6}$ & 30/55 \\
4C +21.35 & J1224.9+2122 & FSRQ & LSP & 0.434 & 194 & $8.6\pm1.5$ & $85.3_{-4.3}^{+4.0}$ & 33/63 \\
PG 1246+586 & J1248.2+5820 & BL Lac & ISP & --- & 373 & $-32.3\pm6.6$ & $86.0_{-3.6}^{+3.9}$ & 45/56 \\
S4 1250+53 & J1253.1+5302 & BL Lac & LSP & --- & 373 & $-16.0\pm7.3$ & $64.3_{-7.7}^{+7.4}$ & 45/52 \\
3C 279 & J1256.1-0547 & FSRQ & LSP & 0.536 & 192 & $19.7\pm3.4$ & $92.3_{-2.9}^{+2.6}$ & 20/70 \\
OP 313 & J1310.6+3222 & FSRQ & LSP & 0.997 & 279 & $6.3\pm7.7$ & $44.1_{-19.1}^{+14.1}$ & 34/58 \\
PKS 1424+240 & J1427.0+2347 & BL Lac & ISP & --- & 386 & $-26.4\pm3.8$ & $68.4_{-9.0}^{+6.9}$ & 36/57 \\
GB6 J1542+6129 & J1542.9+6129 & BL Lac & ISP & --- & 330 & $-3.7\pm6.4$ & $64.0_{-9.3}^{+8.7}$ & 48/57 \\
PKS 1551+130 & J1553.5+1255 & FSRQ & - & 1.308 & 362 & --- & --- & 38/62 \\
PG 1553+113 & J1555.7+1111 & BL Lac & HSP & --- & 421 & $-37.0\pm5.8$ & $89.4_{-1.4}^{+1.5}$ & 37/62 \\
4C +38.41 & J1635.2+3810 & FSRQ & LSP & 1.813 & 199 & $5.9\pm0.8$ & $87.1_{-2.2}^{+1.7}$ & 37/62 \\
S5 1803+784 & J1800.5+7829 & BL Lac & LSP & 0.680 & 269 & $-32.0\pm3.0$ & $57.4_{-28.4}^{+20.5}$ & 17/40 \\
S4 1849+67 & J1849.4+6706 & FSRQ & LSP & 0.657 & 284 & $-18.3\pm2.4$ & $56.9_{-10.3}^{+8.5}$ & 37/60 \\
1ES 1959+650 & J2000.0+6509 & BL Lac & HSP & 0.047 & 439 & $-10.1\pm3.9$ & $50.4_{-9.5}^{+8.6}$ & 23/59 \\
PKS 2144+092 & J2147.3+0930 & FSRQ & LSP & 1.113 & 203 & $-38.6\pm7.3$ & $93.2_{-6.6}^{+4.9}$ & 33/63 \\
BL Lacertae & J2202.8+4216 & BL Lac & ISP & 0.069 & 286 & $-2.6\pm0.7$ & $78.3_{-6.7}^{+5.0}$ & 33/62 \\
PKS 2201+171 & J2203.4+1726 & FSRQ & LSP & 1.076 & 292 & --- & --- & 37/60 \\
PKS 2227$-$08 & J2229.7$-$0832 & FSRQ & LSP & 1.560 & 205 & $-24.5\pm2.2$ & $82.4_{-7.2}^{+5.5}$ & 38/61 \\
CTA 102 & J2232.4+1143 & FSRQ & LSP & 1.037 & 244 & $11.4\pm0.7$ & $92.3_{-2.8}^{+2.0}$ & 30/57 \\
B2 2234+28A & J2236.4+2828 & BL Lac & LSP & 0.795 & 322 & --- & --- & 36/54 \\
PKS 2233$-$148 & J2236.5$-$1431 & BL Lac & LSP & --- & 275 & $30.7\pm1.2$ & $83.9_{-9.2}^{+7.0}$ & 37/57 \\
RGB J2243+203 & J2243.9+2021 & BL Lac & HSP & --- & 543 & $6.2\pm3.9$ & $65.1_{-19.0}^{+13.8}$ & 49/59 \\

\enddata
\tablecomments{Optical and SED classifications, along with redshifts $z$, were obtained from the Second LAT AGN Catalog \citep{2LAC}. 
When the DCF lacked a clear peak, the centroid lag and peak significance are marked ``---''.
In the last column, the approximate KAIT--{\it Fermi}/LAT overlap duration is given, as a percentage of total months covered by the LAT light curve with at least one KAIT data point / as a percentage of total months covered by the LAT light curve {\it after the KAIT campaign began} with at least one KAIT data point.}

\end{deluxetable}

\end{document}